\documentclass[numsec,webpdf,modern,large]{oup-authoring-template}

\usepackage{listings}
\usepackage{float}

\graphicspath{{Fig/}}

\theoremstyle{thmstyleone}%
\theoremstyle{thmstyletwo}%
\theoremstyle{thmstylethree}%

\newcounter{comments}
\newcommand{\red}[1]{\textcolor{black}{#1}}

\journaltitle{Journal}
\DOI{DOI HERE}
\copyrightyear{2026}
\pubyear{2026}
\access{Advance Access Publication Date: Day Month Year}
\appnotes{Paper}

\firstpage{1}

\title[LAFA: Longitudinal Assessment of Function Annotation]{LAFA: A Framework for Reproducible \underline{L}ongitudinal \underline{A}ssessment of Protein \underline{F}unction \underline{A}nnotation Models}

\author[1,2]{An Phan\ORCID{0009-0004-0154-076X}}
\author[3]{Yanli Wang\ORCID{0000-0001-7864-826X}}
\author[4]{Frimpong Boadu}
\author[3]{Jianlin Cheng\ORCID{0000-0003-0305-2853}}
\author[5]{Predrag Radivojac\ORCID{0000-0002-6769-0793}}
\author[2$\ast$]{Iddo Friedberg\ORCID{0000-0002-1789-8000}}

\authormark{Phan et al.}

\address[1]{\orgdiv{Program in Bioinformatics and Computational Biology}, \orgname{Iowa State University}, \orgaddress{\postcode{50011}, \state{Iowa}, \country{United States}}}
\address[2]{\orgdiv{Department of Veterinary Microbiology and Preventive Medicine}, \orgname{Iowa State University}, \orgaddress{\postcode{50011}, \state{Iowa}, \country{United States}}}
\address[3]{\orgdiv{Department of Electrical Engineering and Computer Science}, \orgname{University of Missouri}, \orgaddress{\postcode{65211}, \state{Missouri}, \country{United States}}}
\address[4]{\orgname{Microsoft Corporation}, \orgaddress{\postcode{98052}, \state{Washington}}}

\address[5]{\orgdiv{Khoury College of Computer Sciences}, \orgname{Northeastern University}, \orgaddress{\postcode{02115}, \state{Massachusetts}, \country{United States}}}

\corresp[$\ast$]{Corresponding author: \href{email:idoerg@iastate.edu}{idoerg@iastate.edu}}

\received{Date}{0}{Year}
\revised{Date}{0}{Year}
\accepted{Date}{0}{Year}

\makeatletter
\AtBeginDocument{%
  \def\ps@opening{%
    \def\@oddfoot{\hfill\small\sffamilyfontbold\thepage}%
    \def\@evenfoot{\small\sffamilyfontbold\thepage\hfill}%
    \let\@oddhead\relax
    \let\@evenhead\relax
  }%
}
\makeatother

\usepackage{etoolbox}
\makeatletter
\patchcmd{\@maketitle}%
  {\hbox to \textwidth{\raisebox{5pt}[0pt]{%
          \parbox[b]{416pt}{\raggedright{{\sffamilyfontcnitalic\fontsize{8bp}{12}\selectfont \@journaltitle}, \sffamilyfontcn\@copyrightyear, \thepage--\thelastpage}\\[1pt]
          {\sffamilyfontcn doi: \@DOI}\\[1pt]
          {\ifx\@access\@empty
          \else
          {\sffamilyfontcn \@access}\fi}
          \vskip1pt
          {\ifx\@appnotes\@empty
          \else
          {\sffamilyfontcn \@appnotes}\fi}
            }}%
          \hfill
          {\color{black!20}\rule{45pt}{55pt}}
          }}%
  {}%
  {}{}%
\makeatother
\AtBeginDocument{%
  \def\ps@opening{%
    \def\@oddfoot{\hfill\small\sffamilyfontbold\thepage}%
    \def\@evenfoot{\small\sffamilyfontbold\thepage\hfill}%
    \let\@oddhead\relax
    \let\@evenhead\relax
  }%
}
\begin{document}

\abstract{
\textbf{Motivation:} Protein function prediction is a challenging task and an open problem in computational biology. The Critical Assessment of protein Function Annotation (CAFA) is a triennial, community-driven initiative that provides an independent, large-scale evaluation of computational methods for protein function prediction through time-delayed benchmarking experiments. CAFA has played a key role in highlighting high-performing methodologies and fostering detailed analysis and exchange of ideas. However, outside the periodic CAFA challenges, there is no platform for the continuous evaluation of newly developed methods and tracking performance as function annotations accumulate. \\
\textbf{Results:} Here we introduce the Longitudinal Assessment of Protein Function Annotation Models server (LAFA) as a persistent benchmarking system for protein function prediction methods. LAFA provides a continuous evaluation of containerized function prediction methods, enabling up-to-date and robust comparative assessment of method performance under evolving ground truth. LAFA accelerates methodological iteration, supports reproducibility, and offers a more dynamic and fine-grained view of progress in protein function prediction.  \\
\textbf{Code and Data Availability:} LAFA is available at \url{https://functionbench.net/}. Detailed evaluation results can be found at \url{https://github.com/anphan0828/CAFA_forever}.\\
\textbf{Contact:} \href{idoerg@iastate.edu}{[ahphan\textbar idoerg]@iastate.edu}}



\maketitle
\section{Introduction}\label{sec1}

Protein function prediction is a challenging task and an open problem in computational biology~\citep{bonetta2020machine
}. There are many methods that are used to computationally predict protein function, and it is important that researchers understand their strengths and weaknesses. Furthermore, developers of such methods, as well as the scientific community at large, need a way to assess how well these methods perform. 
The Critical Assessment of Function Annotation (CAFA) of proteins is the premier computational assessment challenge in the field of protein function prediction. CAFA is a triennial challenge that evaluates protein function prediction models in a prospective manner~\citep{radivojac2013large,zhou2019cafa}. In each CAFA round, a subset of all available protein sequences is selected for prediction targets. After that, protein function prediction methods assign functions to these targets given known protein sequences. Functions are represented using the Gene Ontology (GO), a universal controlled vocabulary of gene and gene product attributes~\cite{gene2026gene}. An annotation accumulation period follows the prediction period, in which protein annotations are manually validated by protein biocurators. After the annotation accumulation period, prediction methods are evaluated against the novel set of labels accumulated, and their performance can be compared using standard measures, such as precision, recall, $F_1$-score, as well as new semantic distance metrics~\citep{piovesan2024cafa,radivojac2013large}. However, the initial annotation accumulation period is typically only four to six months long, after which annotations continue to accumulate, but methods are no longer evaluated continuously, but every three years. It follows that some methods may have been misevaluated if their correctly predicted labels were not validated during the annotation accumulation period, a problem known as the Open World assumption~\citep{dessimoz2013cafa}. In addition, as new annotations accumulate, the training set becomes outdated and no longer reflects the most recent annotations, leading to performance decay for methods that rely on old training data. For example, GO terms can become obsolete (with or without direct replacement), or previously known functions get removed due to incorrect GO annotations~\citep{gene2026gene,schnoes2009annotation}. While the evaluation process has been standardized through the development of the CAFA-evaluator package, most of the data collected before the evaluation step was not robustly maintained in a public repository~\citep{piovesan2024cafa}. Finally, many methods that participate in CAFA are not readily usable, as they are difficult to install or available only as a web server.

To address these issues, we have developed the Longitudinal Assessment of Function Annotation server, or LAFA. LAFA hosts containerized, open-source function prediction methods and continuously evaluates them, providing an up-to-date comparative assessment of the hosted methods. Thus, LAFA is designed to compare methods over much more finely resolved time periods rather than every three years, and also to ensure that submitted methods remain usable and reproducible over time.

The rest of this paper is structured as follows. We first describe LAFA's overall architecture and timeline; we then detail the back-end workflows and front-end features that support continuous evaluation and interactive result exploration. Finally, we conclude by outlining some future directions for LAFA as a continuous, community-driven benchmarking system for protein function prediction, and as a system that promotes openness and reproducibility of function prediction methods. We call upon developers of protein function prediction methods to host their methods on the LAFA server.

\section{Implementation}
In CAFA, protein function prediction methods are evaluated through a prospective, time-stratified experimental design. Target proteins are released at a specified time point ($t_{\text{release}}$), predictions are collected by a submission deadline ($t_0$), and performance is evaluated against ground truth annotations accumulated during a subsequent accumulation period (ending at $t_1$). This framework ensures that ground truth annotations used for evaluation did not exist during model training, thereby reducing data leakage and providing a realistic assessment of predictive performance.

LAFA is a benchmarking server that hosts methods for protein function prediction and continuously evaluates them with each new data release. In LAFA, we merge $t_{\text{release}}$ and $t_0$ to be $t_0$, at which targets are specified, and predictions are immediately produced. The development period is no longer needed here since methods participating in LAFA have already been trained and containerized prior to being hosted on the server. Containerization is a software deployment method that packages an application's code, dependencies, and configuration into a single, portable unit~\citep{moreau2023containers}. Hosting methods as containers enables reliable and reproducible results, and are supported widely on different operating systems and computing infrastructure. Specifically for LAFA, containerization also mitigates leakage by preventing the assessed methods from accessing data after the designated $t_0$ date.

\subsection{LAFA infrastructure}
The LAFA infrastructure includes two distinct components: a back-end for computation and a front-end for data visualization. When containerized methods are submitted, we will check their documentation and usability before integrating them into the back-end. The back-end (hosted on a high-performance computing cluster) handles all computational processes including data processing, generating predictions, and evaluating them. The front-end collects updated results from the back-end, and provides a website for browsing, visualizing, and comparing all published LAFA evaluations (see Figure S1). 

The back-end component is currently implemented in Nextflow, a scientific workflow system predominantly used for bioinformatics data analysis~\citep{langer2025empowering,di2017nextflow}. Helper scripts for CAFA and LAFA used in the Nextflow workflow are publicly available at \url{https://github.com/anphan0828/democafa_package}. Back-end and front-end components for LAFA are available at \url{https://github.com/anphan0828/CAFA_forever}. Containerization instructions for hosted methods are available at~\url{https://github.com/anphan0828/LAFA_container_guide}.

\begin{figure*}[tb]
    \centering
    \includegraphics[alt={Timeline diagram of the LAFA evaluation framework spanning May 2025 to March 2026. Horizontal bars indicate model development phases: Methods A and B are trained starting around May 2025, Method C training begins near September 2025, and Method A is retrained to produce A1 around November 2025. An “Annotation Accumulation” bar extends from approximately September 2025 through March 2026, indicating continuous growth of labeled data used for evaluation. At specific data release time points (September 2025, November 2025, December 2025, and March 2026), boxes above the timeline show repeated steps: “Collect data” and “Generate predictions.” These steps produce predictions from the methods available at that time, labeled as A, B, C, and later A1. Colored boxes denote the contributing methods, with green representing A, B, and C, and purple representing the retrained A1. Curved arrows below the timeline represent evaluation windows between consecutive releases. Earlier windows (e.g., September–November 2025) evaluate only methods available at the start of the interval (A and B), while later windows include additional methods as they become available. A highlighted note indicates a longer evaluation window for A and B. The design emphasizes longitudinal comparison and assessment of retraining effects.}, width=\linewidth]{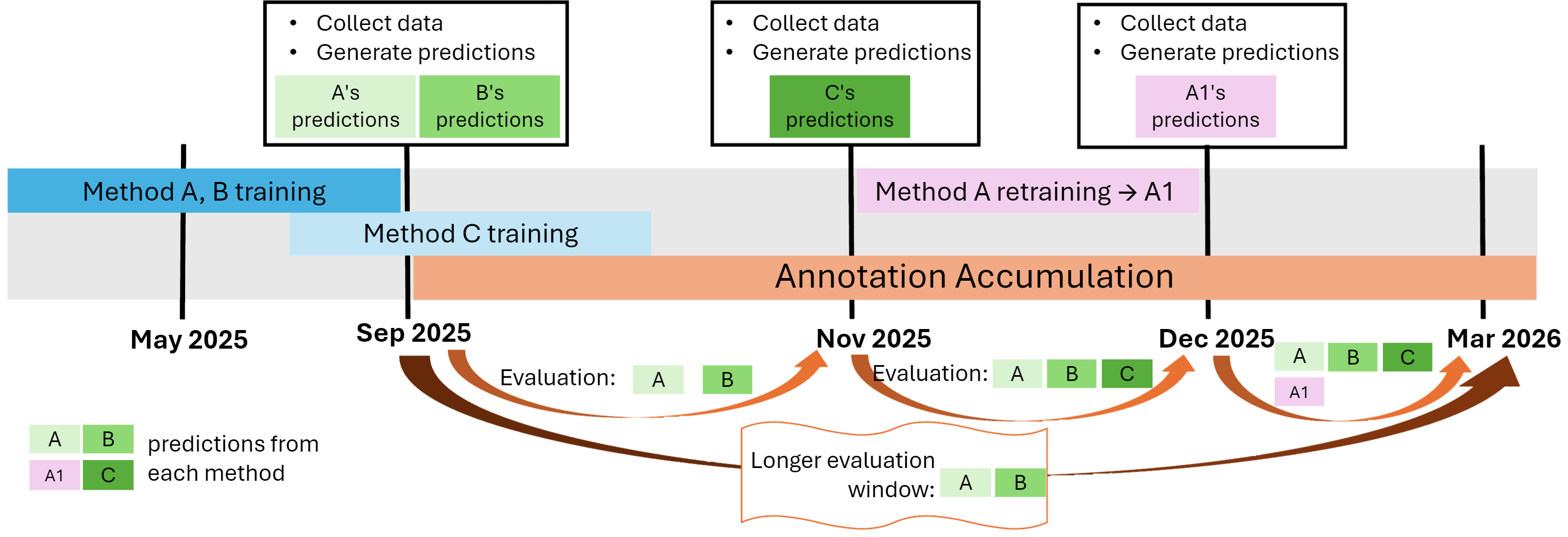}
    
    \caption{\small{\textbf{The LAFA timeline}. At each data release time point (here, starting Sep 2025), data are collected, and predictions are generated from the hosted methods. A time window is defined between any two time points (e.g., Sep 2025 - Nov 2025), during which an evaluation is performed for predictions that existed at the earlier time point (here, only method A and B are included in the Sep 2025 - Nov 2025 evaluation). Evaluation of each time window is accessible on the LAFA website, with an option to compare between any two windows. If a participating method retrains their model with updated training data and submits a new container (here, method A gets retrained to A1), we will generate new predictions and add them to the next evaluation round. The retrained method will then be comparable with its previous version to interpret the effect of training data on evaluation.}}
    \label{fig:workflow}
\end{figure*}



\subsection{Back-end}
All of LAFA's computational assessments occur on the back-end, which is hosted on a high-performance computing system, including: (1) data acquisition and preprocessing, (2) generation of predictions from containerized methods, and (3) evaluation of predictions for each time period. The full implementation of LAFA's backend can be broken down into two main workflows: time point build and time window evaluation, each addressing a distinct task of longitudinal function annotation analysis. Together, these workflows support reproducible construction of a historical protein annotation snapshot, quantitative evaluation across two time points, and post hoc incorporation of late-arriving prediction sets. An example timeline for the evaluation of three methods in LAFA is described in Figure~\ref{fig:workflow}. 

\subsubsection{Time point build workflow}
The time point build workflow is used to construct a complete bundle of processed data (filtered GO annotations, prediction files) from upstream external resources. 
Figure~\ref{fig:workflow} shows several time points from May 2025 to Mar 2026, each of which is linked to a UniProt-GOA data release (approximately once every eight weeks). At each time point, we use the time-point build workflow described below. 

External raw data are acquired from several sources: UniProt-GOA, UniProt Knowledgebase, and the GO Data Archive. UniProt-GOA contains GO annotations of proteins in UniProt, the UniProt Knowledgebase contains protein sequences \red{and GO annotations in SwissProt DAT file (for datasets created after March 2026)}, and GO Data Archive hosts the structure of GO graph. The periodic version release time of UniProt-GOA and UniProt Knowledgebase do not overlap; therefore, we obtain the protein sequences \red{and GO annotations} available at the time of the UniProt-GOA data release. We keep only proteins in SwissProt to constitute a \texttt{test set}. SwissProt is a high-quality, manually annotated, and non-redundant protein sequence database, which constitutes the reviewed section of the UniProt Knowledgebase~\citep{uniprot2025uniprot}. The \texttt{test set} is a relatively stable set of proteins on which prediction algorithms can be compared between time points. In addition, we also construct a \texttt{training set} from proteins with experimentally validated GO annotations. The \texttt{training set} reflects the current high-quality knowledge of protein functions, and annotations of proteins in this set can be transferred to other proteins with little to no GO annotations. \red{For the testbed stage of LAFA, we used a small \texttt{test set} of proteins containing 7,401 proteins with unaltered amino acid sequences that have already accumulated experimental annotations between the 9/2025 and 3/2026 UniProt-GOA releases (see Table~\ref{tab:data_versions} for all time points and time periods). The \texttt{test set} will be expanded to include all proteins in SwissProt, ranging from 550,000 to 580,000 protein sequences in the last ten years, in the future.}

The processed data (training and test sets) are then published on a HuggingFace data repository, making it easier for future participants to test their methods before submitting to LAFA hosting. We also provide a guide to containerizing methods with required input and output for LAFA~\url{https://github.com/anphan0828/LAFA_container_guide}.

After data preprocessing, we begin generating predictions on the \texttt{test set} using existing methods. Registered LAFA participants' containerized methods are pulled from Dockerhub into our high-performance computing system. This happens only once per containerized method. We currently have three containerized methods participating in LAFA, in addition to four baseline predictors.

\begin{itemize}
    \item \textbf{TransFew}: combines function-relevant representations of proteins and semantic representations of GO terms. TransFew was trained on GO annotations in November 2022 and tested on December 2023 data~\citep{boadu2024improving_transfew}.
    \item \textbf{FunBind}: combines protein sequences, textual descriptions, domain annotations, structural features, and GO terms into a pre-trained multimodal foundational model. FunBind was trained on GO annotations in November 2022 and tested on data from December 2024. The version implemented on LAFA only uses the sequence modality~\citep{boadu2025unified_funbind}.
    \item \textbf{DeepGOPlus}: combines learned features from deep convolutional neural networks and sequence similarity-based predictions. The containerized version of DeepGOPlus was trained and tested on GO annotations in May 2025 and protein sequences in June 2025~\citep{kulmanov2020deepgoplus}.
    \item \textbf{Naive baseline}: assigns every protein with every GO term with a score equal to the term's frequency in the database. 
    \item \textbf{Non-Experimental GOA baseline}: predicts terms with a score of 1.0 if the term was annotated at $t_0$. This baseline shows how well non-experimentally validated annotations predict future experimentally validated annotations.
    \item \textbf{BLAST baseline}: transfers GO annotations from homologous proteins via BLASTp search, with scores derived from the highest sequence similarity among matching annotated proteins.
    \item \textbf{Embedding similarity baseline}: transfers annotations from proteins with similar learned representations (from the ProtT5 protein language model), replacing sequence similarity with embedding-space similarity while using the same reference database compared to BLAST baseline~\citep{9477085}.
\end{itemize}

\red{The four baselines, which were also used in previous rounds of CAFA, represent simple approaches to protein function prediction, providing a broad and informative set of reference points for evaluating method performance (see Supplementary Materials for more details).} All seven existing containers are offline: once the required external data has been downloaded, no new data can be accessed, and no connectivity is needed to run the methods.
Predictions are then generated for all proteins in the \texttt{test set}, with the option to split the set into smaller, parallel batches for methods that require more compute resources.
We generated predictions retrospectively on a smaller set of proteins for the testbed stage of LAFA. \red{As mentioned previously, t}his set of proteins, called \texttt{ground truth target set} here, contains 7,401 proteins with unaltered amino acid sequences that have already accumulated experimental annotations between the 9/2025 and 3/2026 UniProt-GOA releases. This accumulation window ensures that all methods only use data available before ground truth annotations are accumulated. Note that the small \texttt{ground truth target set} is used only for analysis in this study, and we will scale LAFA to the full extent of SwissProt as compute resources become available. After the prediction step, large prediction files are stored in each $t_0$ folder and are not published to the front-end.



It is worth noting that TransFew, FunBind, and DeepGOPlus are already trained methods, and take protein sequences as the only required input. This means that as long as the input protein sequences remain unaltered, predictions from these methods are unchanged. Predictions for the same method are regenerated only when the training data for that method is updated, and the retrained method is uploaded to LAFA (see example: method A1 in Figure~\ref{fig:workflow}).

\begin{table}[tb]%
\begin{minipage}{\columnwidth}
\caption{Versions and release dates of external data used in LAFA.}
\label{tab:data_versions}
\begin{tabular}{@{}lcccc@{}}%
\toprule
Release ID & Sep 2025 & Nov 2025 & Dec 2025 & Mar 2026\\
\midrule
UniProt-GOA &  
2025-09-04 & 2025-11-10 & 2025-12-04 & \red{2026-03-04} \\
GO graph date & 2025-07-22 & 2025-10-10 & 2025-10-10 & 2026-01-23 \\
UniProt date & 2025-06-18 & 2025-10-15 & 2025-10-15 & 2026-01-28 \\
\botrule
\end{tabular}
\end{minipage}
\end{table}

\subsubsection{Time window evaluation workflow}
The time window evaluation workflow is used to compute the performance of methods for a given time window. A time window is created between a new time point $t_1$ (when a new data release is available) and any previous time point. Figure~\ref{fig:workflow} displays the example time windows as arrows underneath the time points. The time window evaluation workflow collects a new ground truth (e.g., from Sep 2025 to Nov 2025), and evaluates predictions available at $t_0$ (e.g., Sep 2025) as described below.

First, a common test set is created that contains protein sequences that remain unchanged during this time window. Second, a ground truth is constructed from experimentally validated annotations of proteins in the common test set that accumulated between $t_0$ and $t_1$. 
Third, predictions are evaluated with the CAFA-evaluator package using the obtained ground truth: the predictions (in the format of protein-term-score) are compared with the ground truth (in the format of protein-term) by treating the prediction task as a collection of binary predictions~\citep{piovesan2024cafa}. \red{All protein-term pairs in the predictions and the ground truth are propagated to the GO aspect roots (molecular function, biological process, and cellular component) before evaluation. The rationale for propagating is to consider partially correct predictions (e.g., when a predicted GO term is more specific than the ground truth), allowing all levels of the GO predictions to be evaluated. Additionally, we also assign weights to different levels of GO terms in the graph to capture the specificity of the predictions.} Precision, recall, and $F_1$ measures are computed for each threshold (with a threshold step size of $0.01$ over the score of predictions), and the $F_1$ score at the best threshold is reported as the overall performance metric (see the Supplementary Materials for more details on the evaluation metrics). Finally, completed evaluation results are assembled into release folders named $t_0$\_$t_1$, and a catalog file is updated to indicate the status of all ground truth time windows. Only time windows with status ``ready'' are copied to the official release folder.

Specific instructions on running all LAFA's workflows can be found in the Supplementary Materials.

\subsection{Front-end}
LAFA's front-end is a lightweight website that provides interactive visualization of every time window evaluation. No computation is performed on the front-end, except for rendering plots and selecting methods and time points. The front-end website is packaged inside a Docker container and can be deployed from any web server (currently, we are using an AWS EC2 instance). When a new time window completes the back-end computation, it becomes ``ready'' in the release folder, and only time windows with ``ready'' status are used on the front-end to ensure stable website performance. The Docker image of the website, containing new updates, is then generated and the website is deployed to reflect the most up-to-date results.

The LAFA website provides information about data releases and displays evaluation results continuously across available time points/time windows. The number of targets per time period and the overall performance of all methods help users quickly grasp the performance of protein function prediction methods. Fine-grained evaluation results for each GO aspect -- molecular function, biological process, and cellular component -- are also available in two formats: an $F_1$ score at the best threshold, or precision-recall curves across all thresholds. 

As a benchmarking system, LAFA offers the option to compare evaluations across two selected time windows to track method performance over time. For example, we can compare the performance of methods and baselines between a 4-month and an 8-month annotation accumulation window. Additionally, since LAFA hosts containerized methods that can generate predictions for any time window, we can carry out experiments on the effect of training data recency. In particular, when a method updates its training data and submits a new version, we can compare the performance of the same method when the training data is recent versus outdated on a similar annotation accumulation window. Both use cases are prominent in helping the community answer existing and upcoming questions about the robustness of evaluation and the performance decay of protein function prediction methods over time.


\section{Discussion}
We developed a robust, reproducible, open-source preprocessing and evaluation infrastructure for LAFA, a CAFA-style longitudinal protein function prediction benchmark. 
In particular, LAFA's back-end is organized around two reusable workflows so that any past analysis can be reliably reconstructed and any future time point/time window can be added with minimal manual intervention. Furthermore, LAFA is designed as an open benchmarking platform that supports and invites community-contributed evaluation modules (e.g., new metrics, new visualization features), in the spirit of \texttt{nf-core} community-curated workflow ecosystems~\citep{langer2025empowering}.

 

Future work on LAFA should improve both immediacy and interpretability of the current infrastructure. First, collaborating with UniProt-GOA to secure a private hold-out set of newly curated annotations would enable immediate evaluation of new-arriving methods without compromising the time-delayed principle. Second, scaling compute to handle the prediction tasks for all reviewed SwissProt entries ($\approx 550,000$ sequences) will provide a larger benchmark and, consequently, a better assessment of the hosted model's capabilities.  
Third, adding assessment metrics will make longitudinal performance trends more interpretable and less sensitive to ontology artifacts. For example, GO-slim-based evaluation can provide a stable, high-level view of protein function, complementing the existing metrics that better emphasize GO term specificity and allow partial credits for predicting ancestral terms.

In summary, LAFA extends the CAFA paradigm from a periodic challenge to a persistent community benchmarking resource for protein function prediction. By hosting containerized, open-source methods and evaluating them under temporally consistent conditions, LAFA enables fairer comparison and more transparent measurement of progress over time. We expect that this framework will help provide a foundation for broader community participation in defining how progress in protein function prediction should be measured.





\subsection{For model developers}

If you wish to host your model on LAFA, please contact the corresponding authors. Containerization instructions are available here: \url{https://github.com/anphan0828/LAFA_container_guide} .

\section{Acknowledgments}
\red{The authors thank Maxat Kulmanov and Robert Hoehndorf for contributing DeepGOPlus to the LAFA server. We also thank Henri Chung for useful comments, design suggestions, and documentation suggestions for the website.} 

This work is supported in part by the National Institute of Health [R01GM145937 to IF]; \red{and} the National Science Foundation [DBI2308699 to JC].

\bibliographystyle{abbrvnat}
\bibliography{reference}







\end{document}